\documentclass[preprintnumbers,amsmath,amssymb]{revtex4}
\usepackage{graphicx}
\usepackage{makeidx}
\usepackage{amssymb}
\usepackage{subfigure}
\usepackage{chemarr}
\usepackage{bm}

\def\epsilon{\varepsilon}

\def\beqr{\begin{eqnarray}}
\def\eqnr{\end{eqnarray}}
\def\beq{\begin{equation}}
\def\bc{\begin{center}}
\def\ec{\end{center}}
\def\eqn{\end{equation}}

\begin{document}
\title{Synchronization of two R\"ossler systems with switching coupling}
\author{Mattia Frasca$^1$, Arturo Buscarino$^1$, Marco Branciforte$^2$, Luigi Fortuna$^1$, and Julien Clinton Sprott$^3$}
\affiliation{$^1$Dipartimento
di Ingegneria Elettrica Elettronica e Informatica, Universit\`a degli Studi di Catania, viale A. Doria 6,
95125 Catania, Italy\\ $^2$ST Microelectronics, Stradale Primosole 50, Catania (CT) 95121, Italy\\$^3$Department of Physics, University of Wisconsin - Madison,\\ Madison, WI 53706, USA}

\begin{abstract}
In this paper, we study a system of two R\"ossler oscillators coupled through a time-varying link, periodically switching between two values. We analyze the system behavior with respect to the frequency of the switching. By applying an averaging technique under the hypothesis of a high switching frequency, we find that although each value of the coupling does not produce synchronization, switching between the two at a high frequency stabilizes the synchronization manifold. However, we also find windows of synchronization below the value predicted by this technique, and we develop a master stability function to explain the appearance of these windows. Spectral properties of the system are a useful tool for understanding the dynamical properties and the synchronization failure in some intervals of the switching frequency. Numerical and experimental results in agreement with the analysis are presented.
\end{abstract}


\maketitle


\textbf{}

\section{Introduction}

The interaction of two nonlinear (in particular, chaotic) systems via coupling typically leads to a variety of significant behaviors, among which the most intriguing is probably synchronization, that is, the coordination of a particular dynamical property of their motion \cite{boccaletti02}. The strength and inevitable delay in the coupling affect the type of coordinated dynamical property, leading to different types of synchronization. The most common forms are complete synchronization, i.e. a state in which the two dynamical units evolve following exactly the same trajectory, phase synchronization \cite{rosenblum96,rosa98} when the coordinated property is the phase, lag synchronization \cite{rosenblum97} when both amplitudes and phases are locked but with a permanent time lag, and generalized synchronization \cite{rulkov95} when a given function of the output of two systems is synchronized.

The phenomenon is even more varied when more than two units interact according to a pattern of connectivity through which they share information about their current state: complete synchronization \cite{gomez07}, phase synchronization \cite{arenas08}, cluster synchronization \cite{pecora14}, partial synchronization \cite{wick14}, chimera states \cite{abrams04,gambuzza14}, relay synchronization \cite{gutierrez13}, and remote synchronization \cite{gambuzza13} are only examples of the behaviors observed in a network of coupled oscillators. For complete synchronization as investigated in this paper, a series of well-known results provides the conditions for the onset of the synchronous motion. If the synchronization manifold is a stable solution for the given network, a suitable coupling configuration can always be designed \cite{msf}.

Although the connectivity among the dynamical units is usually considered time-invariant, interaction among dynamical systems may also occur in a discontinuous way (for instance, when it is mediated by links activated according to the relative distance of mobile units \cite{frasca08,frasca12,fujiwara11}) or with weights varying in time according to some adaptation law \cite{yu12,lehnert14,gambuzza15}. In such cases, a key factor in determining the global behavior of the system is the interplay between its time scales: one related to dynamics of the units, and the other defining the rate of variation of the links between them.

Another important ingredient is the dynamical rule for the variation of the coupling, which may be either a stochastic/periodic activation/deactivation \cite{belykh04,buscarino15} or a deterministic law \cite{hasler13}. Both possibilities were explored in early experimental works on two coupled Chua's circuits \cite{Chua96,Fortuna2003}. In particular, in \cite{Chua96} adaptive coupling was used to design communication channels able to compensate for parameter changes, while in \cite{Fortuna2003} it was demonstrated that in a synchronization scheme where two Chua's circuits are pulse coupled, the switching frequency has to be larger than a threshold value to achieve synchronization. This behavior is now grounded on recent theoretical results for blinking networks \cite{skufca04,porfiri06}, proving that under an assumption referred to as the fast switching hypothesis (FSH), when the link changes occur \textit{enough} faster than the oscillator dynamics, the time-varying coupling can be studied by means of the time-average of the coupling matrix.

However, since there is some arbitrariness in the definition of ``\textit{fast enough},'' procedures to determine explicit bounds for the time-scale of the process driving the coupling mechanism are currently under investigation \cite{hasler13}. On the other hand, there is evidence that even below the threshold given by the FSH, many interesting phenomena may occur. For instance, a recent study on synchronization of chaotic oscillators coupled via an on-off stochastic network has unveiled non-trivial windows of complete synchronization even when, under the FSH, synchronization is not predicted \cite{jeter15}.

This paper is a case study of two R\"ossler systems interacting through on-off coupling and showing interesting phenomena in a regime not dominated by fast switching. We consider the case where the two chaotic systems are coupled through a link whose weight is time-varying, i.e. the weight switches between two fixed values with a given switching frequency. We fix the values so that neither of them provides synchronization in the case of static coupling. Despite this, switching between the two values gives synchronous behavior. The stability of the synchronization manifold is studied for switching frequencies spanning the slow to fast switching regimes, unveiling the relationship between the two time-scales. We also provide experimental validation of the results using a hybrid platform where two analog electrical circuits are coupled by digitally controlled links.

The rest of the paper is organized as follows: in Section~\ref{sec:model} the model is described; in Section~\ref{sec:results} the numerical results are shown, and in Section~\ref{sec:analysis} the analysis with respect to the switching frequency is presented, while Section~\ref{sec:expresults} deals with the experimental validation of the results. Section~\ref{sec:conclusions} gives conclusions.

\section{Model}
\label{sec:model}

The R\"ossler oscillator is described by the following nonlinear dynamical equations \cite{rossler76}:

\begin{equation}
\begin{array}{l}
\dot{x}=-y-z\\
\dot{y}=x+ay\\
\dot{z}=b+z(x-c)
\end{array}
\label{eq:rossler}
\end{equation}

\noindent where $a$, $b$, and $c$ are system parameters. We fix $a=b=0.2$ and $c=7$ throughout the rest of the paper so that a chaotic attractor is obtained in the range of considered initial conditions.

We first consider two identical R\"ossler systems with a time-invariant diffusive coupling acting between the first components of their state space, and we briefly discuss the behavior of this configuration as a prelude to the fast switching approach. The equations governing the two coupled systems can be written as

\begin{equation}
\begin{array}{l}
\dot{x}_1=-y_1-z_1+\kappa(x_2-x_1)\\
\dot{y}_1=x_1+ay_1\\
\dot{z}_1=b+z_1(x_1-c)\\
\dot{x}_2=-y_2-z_2+\kappa(x_1-x_2)\\
\dot{y}_2=x_2+ay_2\\
\dot{z}_2=b+z_2(x_2-c)
\end{array}
\label{eq:rosslerCoupl}
\end{equation}

\noindent where the subscripts indicate the two respective systems, and $\kappa$ is the coupling strength. For brevity, we define $\textbf{x}_1=[\begin{array}{lll} x_1 & y_1 & z_1 \end{array}]^T$ and $\textbf{x}_2=[\begin{array}{lll} x_2 & y_2 & z_2 \end{array}]^T$, respectively, and rewrite Eqs.~(\ref{eq:rosslerCoupl}) compactly as

\begin{equation}
\begin{array}{l}
\dot{\textbf{x}}_1=f({\textbf{x}}_1)+\kappa\mathrm{E}(\textbf{x}_2-\textbf{x}_1)\\
\dot{\textbf{x}}_2=f({\textbf{x}}_2)+\kappa\mathrm{E}(\textbf{x}_1-\textbf{x}_2)
\end{array}
\label{eq:rosslerCouplCompact}
\end{equation}

\noindent where $\mathrm{E}=\left [ \begin{array}{lll} 1 & 0 & 0 \\ 0 & 0 & 0 \\ 0 & 0 & 0 \end{array} \right ]$. Complete synchronization is formally defined as

\begin{equation}
\label{eq:syncdef}
\| \textbf{x}_1 - \textbf{x}_2 \| \rightarrow 0, \texttt{~as~} t \rightarrow \infty
\end{equation}

\noindent where $\| \cdot \|$ stands for the Euclidean norm. To derive the conditions on $\kappa$ for complete synchronization, we define the error as $\textbf{e}(t)=\textbf{x}_1 - \textbf{x}_2$ and calculate the error dynamics from Eqs.~(\ref{eq:rosslerCouplCompact}):

\begin{equation}
\begin{array}{l}
\frac{d(\textbf{x}_1-\textbf{x}_2)}{dt}=f({\textbf{x}}_1)-f({\textbf{x}}_2) - 2\kappa\mathrm{E}(\textbf{x}_1-\textbf{x}_2)
\end{array}
\label{eq:errordynamics}
\end{equation}

By linearizing around the common solution $\textbf{x}_1=\textbf{x}_2=\textbf{s}$, we obtain

\begin{equation}
\begin{array}{l}
\frac{d\textbf{e}}{dt}=\left ( \left. \frac{\partial f(\textbf{x})}{\partial \textbf{x}}\right |_{\textbf{x}=\textbf{s}} - 2\kappa\mathrm{E} \right )\textbf{e}
\end{array}
\label{eq:msfstatic}
\end{equation}

\begin{figure}
\centering
  \includegraphics[height=.22\textheight]{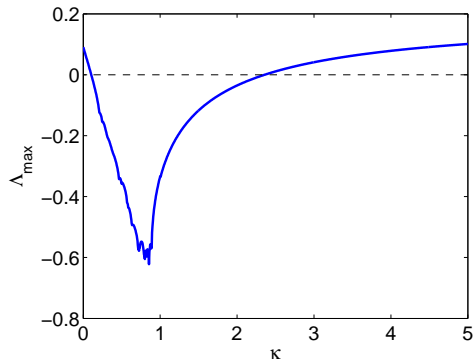}
  \caption{Master Stability Function for the system in Eqs.~(\ref{eq:rosslerCoupl}) with static coupling. \label{fig:msfRoss}}
\end{figure}

The maximum (largest) Lyapunov exponent $\Lambda_{max}(\kappa)$ of Eq.~(\ref{eq:msfstatic}) is a function of $\kappa$ and indicates the region of local stability of the error dynamics, and so of complete synchronization. Following the terminology of \cite{msf}, we refer to it as the Master Stability Function (MSF) of the system. In particular, synchronization requires that $\Lambda_{max}(\kappa)<0$. Figure \ref{fig:msfRoss} displays the MSF for the static coupling case of Eqs.~(\ref{eq:rosslerCoupl}). Note that $\Lambda_{max}(\kappa)<0$ only in the interval $0.1<\kappa<2.35$. This behavior, referred to as a class-II MSF \cite{boccaletti08}, is characteristic of a class of systems, including the R\"ossler oscillator when coupled through the variable $x$.

We now focus on the main object of the study, which is a system formed by two R\"ossler units coupled with a time-varying link given by

\begin{equation}
\begin{array}{l}
\dot{x}_1=-y_1-z_1+\kappa(t)(x_2-x_1)\\
\dot{y}_1=x_1+ay_1\\
\dot{z}_1=b+z_1(x_1-c)\\
\dot{x}_2=-y_2-z_2+\kappa(t)(x_1-x_2)\\
\dot{y}_2=x_2+ay_2\\
\dot{z}_2=b+z_2(x_2-c)
\end{array}
\label{eq:rosslerCouplBlink}
\end{equation}

We assume the coupling is given by $\kappa(t)=k_1+\frac{k_2-k_1}{2}(\text{sgn}(\cos(\omega t))+1)$ where $\text{sgn}(x)=1$ for $x>0$ and $\text{sgn}(x)=-1$ otherwise, so that the effective coupling switches between two constant values $k_1$ and $k_2$ at a frequency $\omega$. We refer to Eqs.~(\ref{eq:rosslerCouplBlink}) as the switching system and analyze its behavior with respect to the switching frequency $\omega$, which is an important bifurcation parameter.

In particular, we select $k_1$ and $k_2$ such that neither of the two falls within the synchronization range for the static coupled case (MSF of Fig.~\ref{fig:msfRoss}), that is, $\Lambda_{max}(k_1)>0$ and $\Lambda_{max}(k_2)>0$. Under these conditions, the problem of synchronization of the switching system is not trivial since the system switches between two configurations that are not synchronizable. A recently developed approach for blinking systems \cite{hasler13} provides a useful tool for understanding the behavior of the system when the switching occurs at a high enough frequency. For an average system defined by substituting for the coupling parameter its average value (in our case $k(t)=\bar{k}=\frac{k_1+k_2}{2}$), the trajectory of the switching system approaches that of the average system under the hypothesis of fast switching. We will show that the fast switching approach can be used to predict the behavior of the switching system, although synchronization can also occur at lower frequencies within windows of $\omega$.

\section{Numerical results}
\label{sec:results}

In the numerical simulations, we set $k_1=0$ and $k_2=2.4$. Both coupling values fall outside the range of synchronization identified by the MSF of Fig.~\ref{fig:msfRoss}, i.e., $\Lambda_{max}(k_1)>0$ and $\Lambda_{max}(k_2)>0$, while the average value $\bar{k}=1.2$ lies in the stable region where $\Lambda_{max}(\bar{k})<0$.

We investigate the effect of the switching frequency by fixing all the other parameters and varying $\omega$ from $0$ to $1.5$ (the limit case $\omega=0$ corresponds to uncoupled dynamics). For each value of $\omega$, we integrate Eqs.~(\ref{eq:rosslerCouplBlink}) with a 4th order adaptive Runge-Kutta algorithm for a time $T=10^7$s and sample the result at $dt=0.01$, thus obtaining $M=10^9$ samples from which the average synchronization error $E(\omega)$ is calculated from

\begin{equation}
E(\omega)=\sum_{h=1}^M\frac{\sqrt{(x_1(h)-x_2(h))^2+(y_1(h)-y_2(h))^2+(z_1(h)-z_2(h))^2}}{\sqrt{x_1(h)^2+y_1(h)^2+z_1(h)^2+x_2(h)^2+y_2(h)^2+z_2(h)^2}}
\label{eq:errsync}
\end{equation}

The synchronization error $E(\omega)$ is normalized so that $E=1$ means that the two trajectories are completely uncorrelated, $E>1$ indicates anti-correlation, while $E\rightarrow 0$ corresponds to the highest correlation.

\begin{figure}
\centering
\includegraphics[height=.22\textheight]{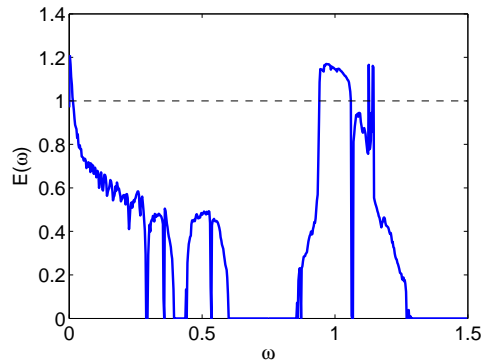}
\caption{Average synchronization error $E(\omega)$ with respect to the switching frequency $\omega$. \label{fig:errNum}}
\end{figure}

\begin{figure}
\centering
  \includegraphics[height=.22\textheight]{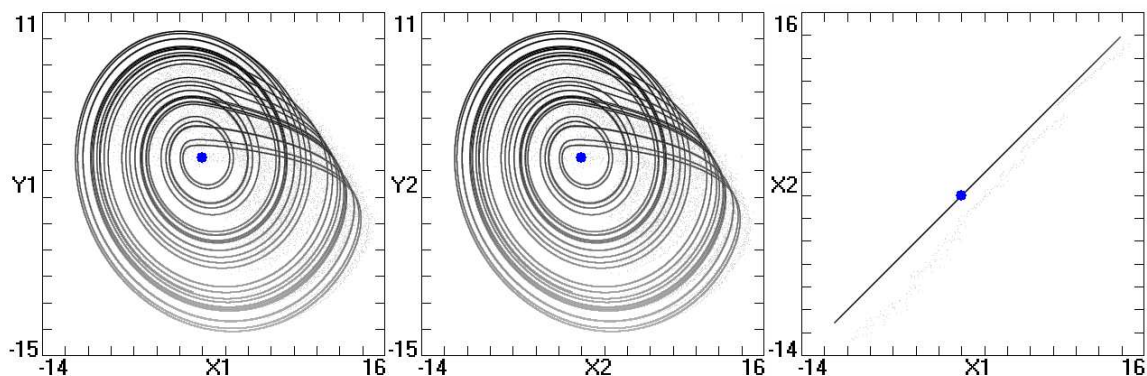}
  \caption{Complete synchronization of two R\"ossler oscillators coupled by a switching signal with $\omega=1.5$. Projection of the attractor onto the planes (a)   $x_1$-$y_1$ , (b) $x_2$-$y_2$, (c) $x_1$-$x_2$. \label{fig:fsh}}
\end{figure}

The synchronization error $E(\omega)$ in Fig.~\ref{fig:errNum} is a nonmonotonic function of $\omega$. In particular, for $\omega>1.3$ the system is synchronized. Therefore, $\omega \simeq 1.3$ represents a boundary between the region in which the FSH holds, and that in which the frequency of the switching is not ``fast enough.'' For $\omega > 1.3$, the time-varying connectivity is sufficiently fast relative to the R\"ossler dynamics that the system responds as it would to a constant coupling equal to the average of the two coupling strengths $k_1$ and $k_2$. In fact, the two oscillators in this regime are always synchronized and chaotic as shown in Fig.~\ref{fig:fsh} for $\omega = 1.5$.

The behavior at low frequencies ($\omega <0.2$) is expected from the choice of $k_1$ and $k_2$. In this case, the system slowly alternates between two configurations, neither of which is synchronizable. The global behavior is thus unsynchronized.

The most interesting frequencies are in the range $\omega \in [0.2, 1.3]$ where nontrivial alternating windows of synchronization and nonsynchronization occur. In particular, even when the FSH does not hold and the oscillating link has time scales comparable to those of the R\"ossler system, there are regions of complete synchronization.

\begin{figure}
\centering
  \includegraphics[height=.22\textheight]{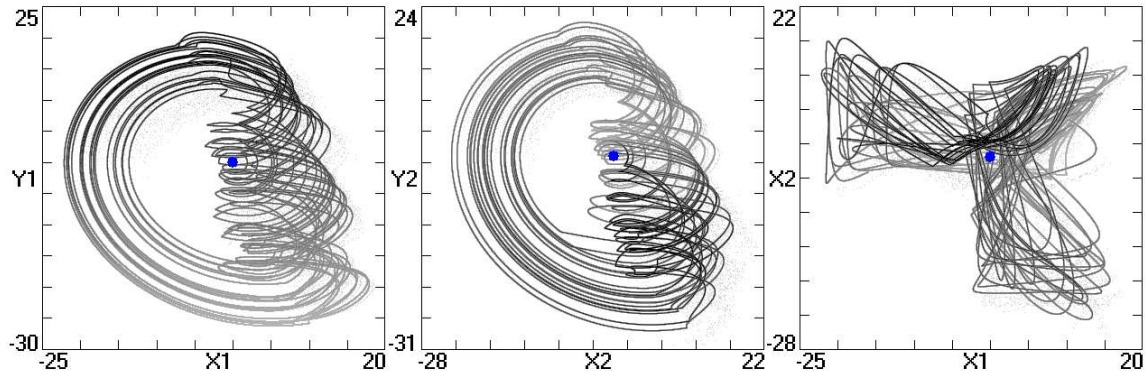}
  \caption{Unsynchronized behavior of two R\"ossler oscillators coupled by a switching signal with $\omega=1$. Projection of the attractor onto the planes (a) $x_1$-$y_1$, (b) $x_2$-$y_2$, (c) $x_1$-$x_2$. \label{fig:peak}}
\end{figure}

Starting from $\omega=1.3$ at the boundary where the FSH holds and decreasing the switching frequency, we examine the dynamical behavior of the R\"ossler oscillators at different values of $\omega$. Just below $\omega=1.3$, a large window ($0.9<\omega<1.3$) of unsynchronized behavior is found. The system attractor is significantly different from the chaotic attractor of the synchronized R\"ossler oscillators as shown in Fig.~\ref{fig:peak} for $\omega=1.0$.

\begin{figure}
\centering
\includegraphics[height=.22\textheight]{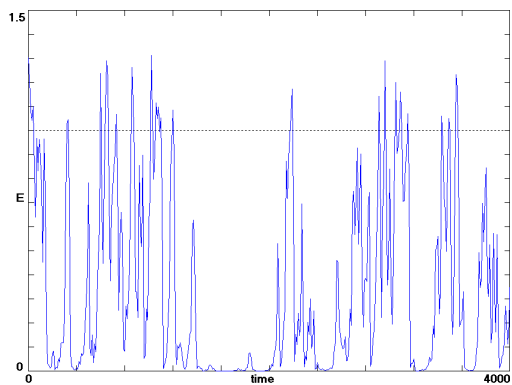}
\caption{Temporal evolution of the average synchronization error $E$ for $\omega=1.13$ showing intermittency. \label{fig:intermittency}}
\end{figure}

\begin{figure}
\centering
  \includegraphics[height=.22\textheight]{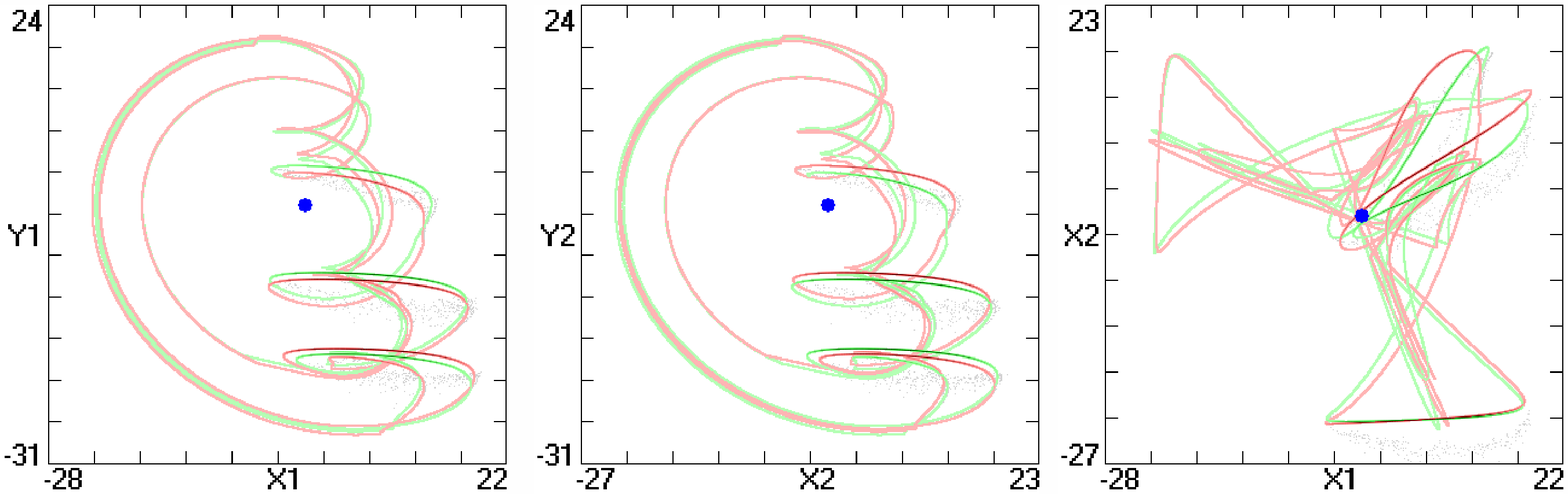}
  \caption{Limit cycles in the two R\"ossler oscillators coupled with a switching signal with $\omega=0.9656$. (a) Projection of the attractor onto the plane $x_1$-$y_1$, (b) projection of the attractor onto the plane $x_2$-$y_2$, and (c) projection of the attractor onto the plane $x_1$-$x_2$. \label{fig:limit09656}}
\end{figure}

We also observe two other significant dynamical behaviors. At $\omega \approx 0.97$ and $\omega \approx 1.13$, which correspond to the transition between synchronous and unsynchronous regions, intermittent synchronization occurs as shown in Fig.~\ref{fig:intermittency}. The shape of the attractor in this case alternates between the original shape of the R\"ossler chaotic attractor (obtained when the synchronization error is close to zero) and the one shown in Fig.~\ref{fig:peak}, obtained when the error is larger. Furthermore, at $\omega \approx 0.9656$ there is a narrow window in which two stable limit cycles coexist as shown in Fig~\ref{fig:limit09656}.

For a further decrease in the switching frequency, a new window of synchronization occurs at $0.6<\omega<0.9$. Below $\omega=0.6$ a series of unsynchronized/synchronized windows are observed whose widths decrease for decreasing values of $\omega$. Within all these windows, including the main one around $\omega=1$, a narrow window of synchronization occurs, only three of which are evident in Fig.~\ref{fig:errNum}.

\begin{figure}
\centering
  \includegraphics[height=.3\textheight]{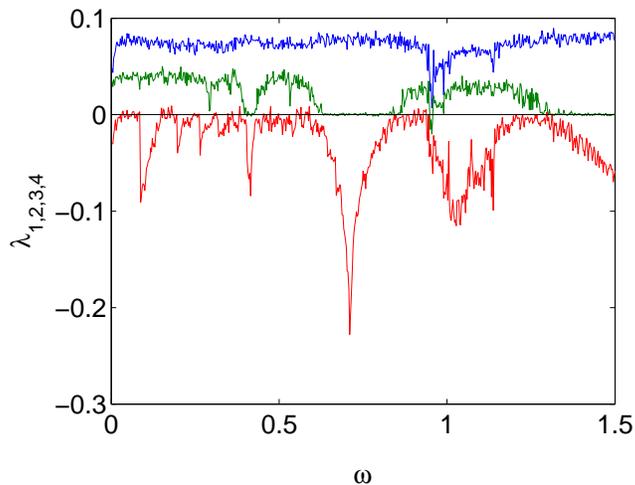}
  \caption{Lyapunov exponent spectrum for the system in Eqs.~(\ref{eq:rosslerCouplBlink}). \label{fig:lyapvsomega}}
\end{figure}

The observed behavior is confirmed by an analysis of the four largest Lyapunov exponents (\ref{eq:rosslerCouplBlink}) shown in Fig.~\ref{fig:lyapvsomega}. Regions of synchronization are characterized by one positive Lyapunov exponent, while those of unsynchronized behavior have two positive Lyapunov exponents corresponding to hyperchaos in the seven-dimensional state space. The sign of the second largest non-zero Lyapunov exponent thus discriminates between the windows of synchronous and unsynchronous motion found by analysis of the synchronization error.

\section{Analysis}
\label{sec:analysis}

From the numerical results of Section~\ref{sec:results}, it is evident that the switching frequency is a bifurcation parameter, especially when the FSH does not hold, that is, when the rate of switching is comparable to the dynamics of the R\"ossler system. The switching frequency affects not only the synchronization but also the dynamics of the attractor. This section provides further analysis of the observed phenomena. In particular, we first investigate the properties of the power spectra for the R\"ossler state variables and correlate them with the windows observed in Figs.~\ref{fig:errNum} and~\ref{fig:lyapvsomega} and then develop a MSF for the case of two oscillators with a time-varying coupling.

\begin{figure}
\centering
  \includegraphics[height=.3\textheight]{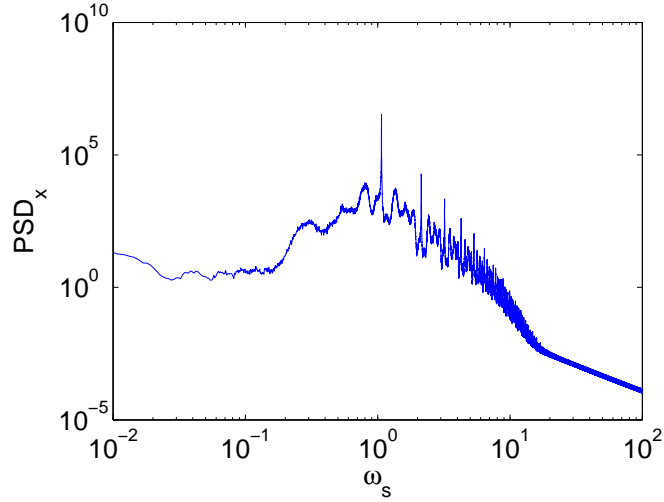}
  \caption{Power spectral density for the $x$ variable of the R\"ossler system in Eq~(\ref{eq:rossler}). \label{fig:power}}
\end{figure}

\begin{figure}
\centering
  \includegraphics[height=.3\textheight]{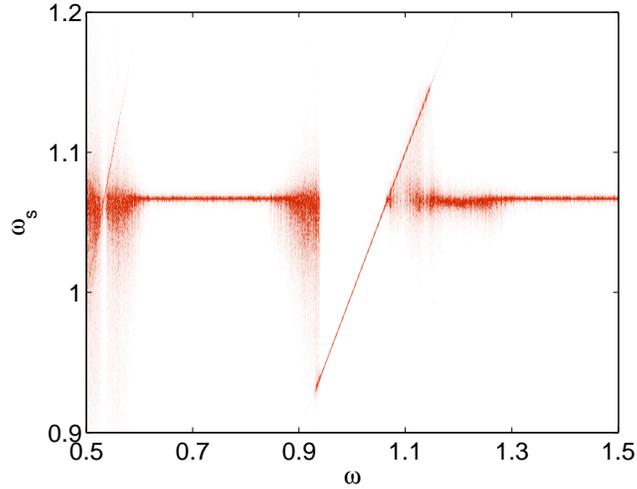}
  \caption{Power spectral density for the $x$ variable of the R\"ossler system in Eq~(\ref{eq:rosslerCouplBlink}), for different values of $\omega$. In the color code red indicates higher power. \label{fig:spectrogram}}
\end{figure}

We begin by observing that the power spectral density of an uncoupled R\"ossler system~(\ref{eq:rossler}) as shown in Fig.~\ref{fig:power} is characterized by a strong dominant component corresponding to the large nearly periodic oscillations in $x$-$y$ near the $z=0$ plane having a frequency $\omega_s \simeq 1.067$. The large window of unsynchronized behavior in Fig.~\ref{fig:errNum} just below the boundary with the fast switching region is around this frequency. A closer inspection of the switching system spectrum shows that a resonance occurs in this window as shown in Fig.~\ref{fig:spectrogram}. The Figure shows the power spectral density (color coded) for $0.9 < \omega_s < 1.2$ as a function of the switching frequency $\omega \in [0.5,1.5]$. In the two windows where the systems are synchronized (one is the fast switching region for $\omega>1.3$, and the other is the window $0.6<\omega<0.9$), the dominant frequency is $\omega_R$, the same as for the uncoupled R\"ossler system. That is, when the two systems synchronize, they evolve following the chaotic trajectory $\textbf{s}$ of the uncoupled system. On the contrary,  when the coupling strength switches at a frequency $\omega$ comparable to $\omega_R$, in particular in the window $0.95 <\omega < 1.15$, the dominant frequency is locked to the switching frequency, thus resulting in a modification of the dominant frequency with respect to the uncoupled case. This explains the different shape of the attractor in this range (Fig.~\ref{fig:peak}). A similar locking occurs at the first subharmonic of the dominant frequency around $\omega=0.5335$ superimposed on the window of unsynchronized behavior. Presumably, the same locking occurs in narrower ranges around all the other subharmonics.

The frequencies of the three unstable periodic orbits with the shortest periods for the single R\"ossler system are $1.0626, 0.5294$, and $0.3533$, corresponding closely to the dominant frequency and its first two subharmonics. Thus it is difficult to distinguish whether the resonances are a result of the peak in the power spectral density or an interaction with the unstable orbits on the attractor. However, it may be significant that the number of unstable periodic orbits proliferate enormously for frequencies below about $\omega = 0.3$, and this may account for the absence of low-frequency structure in $E(\omega)$ shown in Fig.~\ref{fig:errNum}.

\begin{figure}
\centering
  \includegraphics[height=.4\textheight]{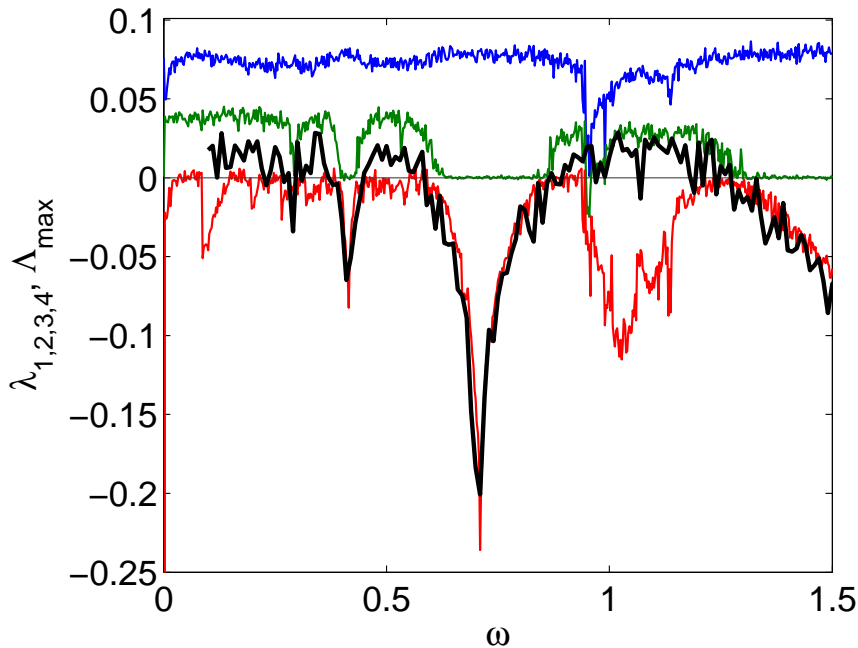}
  \caption{Maximum Lyapunov exponent $\Lambda_{max}(\omega)$  for the system in Eqs.~(\ref{eq:rosslerCouplBlink}) with time-varying coupling. \label{fig:msfRossTV}  }
\end{figure}

We now derive a MSF for the time-varying coupling. To do this, we calculate the error dynamics for system (\ref{eq:rosslerCouplBlink}) and linearize around the common solution $\textbf{x}_1=\textbf{x}_2=\textbf{s}$ to obtain

\begin{equation}
\begin{array}{l}
\frac{d(\textbf{e})}{dt}=\left ( \left. \frac{\partial f(\textbf{x})}{\partial \textbf{x}}\right |_{\textbf{x}=\textbf{s}} - 2\kappa(t)\mathrm{E} \right )\textbf{e}
\end{array}
\label{eq:msftimevarying}
\end{equation}

\noindent with $\kappa(t)=k_1+\frac{k_2-k_1}{2}(\text{sgn}(\cos(\omega t))+1)$. From Eqs.~(\ref{eq:msftimevarying}), the maximum Lyapunov exponent $\Lambda_{max}(\omega)$ as a function of the parameter $\omega$ is calculated as shown in Fig.~\ref{fig:msfRossTV}. Values of $\omega$ such that $\Lambda_{max}(\omega)<0$ correspond to switching signals able to synchronize the R\"ossler oscillators, while $\Lambda_{max}(\omega)>0$ indicates that the error does not decay to zero for that value of the switching frequency. The curve $\Lambda_{max}(\omega)$ vs. $\omega$ perfectly explains the windows of synchronization and nonsynchronization for system (\ref{eq:rosslerCouplBlink}).

\section{Experimental investigation}
\label{sec:expresults}

In this section, the system of two R\"ossler oscillators with switching coupling is experimentally investigated using an electronic circuit governed by Eqs.~(\ref{eq:rosslerCouplBlink}). Each R\"ossler oscillator uses the electrical scheme reported in \cite{buscarinoBook}. The temporal dynamics are rescaled by a factor $K=2128$, so that the circuit waveforms correspond to those of the R\"ossler system (\ref{eq:rossler}) with a rescaled time axis.

Unlike static coupling, which requires only a single resistor between the capacitors associated with the respective coupled state variables, coupling in this experiment used an analog switch in series with a coupling resistor driven by a pulse width modulated (PWM) source. This strategy is implemented in anticipation of extending the studies to a network of switched connections. We have also explored solutions using a digital resistor, but the resolution provided by off-the-shelf digital resistors in terms of values and switching time is inadequate for our purposes.

In more detail, the coupling scheme was implemented by using two components: the analog switch ADG452 and the ST microcontroller unit (MCU) STM32F303VCT6 \cite{st} for the generation and control of the PWM signal. The ADG452 embeds four independently selectable bi-directional switches, has a low on-resistance (on the order of $5\Omega$), fast switching times ($t_{ON}=70 ns$, $t_{OFF}=60 ns$), and is TTL-/CMOS-compatible. The microcontroller STM32F303VCT6 is an ARM\circledR-based Cortex\circledR-M4, 32bit microcontroller with an embedded floating point unit. It has a core clock speed up to $72$MHz, a $256$KB flash memory, $48$KB SRAM, and a wide range of peripherals such as analog-to-digital and digital-to-analog converters, timers, direct memory access, etc. It operates with a voltage supply in the range from $2.0V$ to $3.6V$.

The value of the coupling resistor is controlled by the duty cycle (DC) of the PWM signal driving the analog switch. Turning on and off the switch has the effect of multiplying the fixed coupling resistor by a factor inversely proportional to the DC, according to the equation

\begin{equation}
\label{eq:formula}
R_{eq}=\frac{100}{DC}R_c
\end{equation}


In our application, the ADG452 has been controlled with a $40$kHz PWM, which is a suitable value since the characteristic frequency range of the R\"ossler implementation is below $5$kHz. Although this solution is able to switch between two nonzero values of the coupling resistance, it allows implementing a very low value of the coupling $k_1$ ($k_1=0.02$), which for the purpose of our analysis is equivalent to two disconnected circuits.





The waveforms corresponding to the six state variables generated in the experiment were acquired using an NI-USB6255 data acquisition board at a sampling frequency of $f_s=75kHz$ and post-processed to compute the average normalized error $E(\Omega)$ as in Eq.~(\ref{eq:errsync}), where $\Omega=K\omega$ is the switching frequency in the rescaled circuit.

\begin{figure}
\centering{\includegraphics[height=.22\textheight]{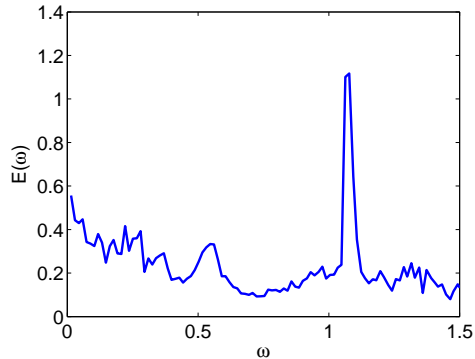}} \caption{Average normalized error $E(\omega)$ calculated on data acquired from the experimental system. Frequency axis is rescaled by the factor $K$ to allow comparison with numerical results. \label{fig:errexp}}
\end{figure}

The trend of $E(\Omega)$ is shown in Fig.~\ref{fig:errexp}. To facilitate the comparison with the numerical results, the frequency has been rescaled in terms of the variable $\omega$. The presence of a main peak of desynchronization around $\omega=1.06$ is also confirmed in the experimental case. Due to component tolerances, in spite of predicted complete synchronization, i.e., $E=0$, only practical synchronization \cite{buscarino09,monten15} ($E<0.2 V$) is observed. The windows of practical synchronization are in good agreement with those of complete synchronization found in the numerical simulations.

\section{Conclusions}
\label{sec:conclusions}

In this paper, two R\"ossler oscillators coupled through a time-varying link have been investigated. In particular, the coupling strength switches periodically between two values. The observed behavior is significantly more rich than the case of other chaotic circuits such as Chua's circuit, for which there is only a single transition in frequency between synchronized and unsynchronized behavior \cite{Fortuna2003}.

Although each of the two values of the coupling, when applied statically to the system, does not lead to synchronization, switching between them can stabilize the synchronization manifold. This behavior is observed not only for high switching frequencies as predicted by the fast switching theory, but also in several windows at lower switching frequencies.

The alternate windows of synchronized and unsynchronized behavior can be explained by a MSF illustrating the behavior of the maximum Lyapunov exponent transverse to the synchronization manifold as a function of the switching frequency. Windows of synchronization correspond to negative values of the maximum transverse Lyapunov exponent, while unsynchronized behavior is obtained when this measure is positive.

We have also demonstrated a strong effect of the switching coupling on the spectral properties of the two oscillators, since a switching frequency very close to the dominant component of the spectrum of the isolated R\"ossler oscillators causes the dominant component to lock to the switching frequency, resulting in a significant change of the chaotic dynamics and a failure of synchronization.

Hence inspection of the spectral properties of the system and analysis of the MSF derived for the time-varying coupling allow one to establish the regions of synchronizability beyond that predicted by the fast switching analysis. These regions agree well with the results of numerical analysis as well as experimental results.

\section{Acknowledgements}
We acknowledge Fabrizio La Rosa for his help in the realization of the experimental setup.

\end{document}